\documentclass[aps,twocolumn,showpacs,amsmath,amssymb,superscriptaddress,prc]{revtex4-2}
\usepackage{graphicx,here}
\usepackage{multirow}
\usepackage{tikz}
\usepackage{epstopdf}
\usepackage[percent]{overpic}
\usepackage{scrextend}
\usepackage{xcolor,xspace}
\usepackage[ulem=normalem]{changes}
\usepackage{hyperref}
\hypersetup{colorlinks=True,urlcolor=blue,linkcolor=blue,citecolor=blue,filecolor=black}

\colorlet{Changes@Color}{red}
\usepackage{dcolumn,color,footnote,bm,braket}
\usepackage{url,longtable,tabularx}
\usepackage{threeparttable}

\usepackage{fancybox}
\usetikzlibrary{matrix}

\begin{document}

\title{Quadrupole-hexadecapole correlations in neutron-rich samarium and gadolinium isotopes}

\author{L. Lotina}
\email{llotina.phy@pmf.hr}
\email{luka.lotina@uam.es}
\affiliation{Department of Physics, Faculty of Science, 
University of Zagreb, HR-10000 Zagreb, Croatia}
\affiliation{Departamento de F\'{i}sica Te\'{o}rica and CIAFF, Universidad Aut\'{o}noma de Madrid, E-28049 Madrid, Spain}

\author{K. Nomura}
\email{nomura@sci.hokudai.ac.jp}
\affiliation{Department of Physics, 
Hokkaido University, Sapporo 060-0810, Japan}
\affiliation{Nuclear Reaction Data Center, 
Hokkaido University, Sapporo 060-0810, Japan}

\author{R. Rodr\'{\i}guez-Guzm\'an}
\email{guzman.rodriguez@nu.edu.kz}
\affiliation{Department of Physics, Nazarbayev 
University, 53 Kabanbay Batyr Ave., Astana 010000, Kazakhstan}

\author{L. M. Robledo}
\affiliation{Departamento de F\'{i}sica Te\'{o}rica and CIAFF, Universidad Aut\'{o}noma de Madrid, E-28049 Madrid, Spain}
\affiliation{Center for Computational Simulation, Universidad Polit\'{e}cnica de Madrid, Campus de Montegancedo, Bohadilla del Monte, E-28660-Madrid, Spain}

\date{\today}

\begin{abstract}
We present an extensive study of quadrupole-hexadecapole 
correlation effects in even-even Sm and Gd isotopes with neutron number $N=88-106$. 
The calculations are performed in the framework 
of the Gogny energy density functional (EDF) 
with the D1S parametrization and 
the $sdg$ interacting boson model (IBM). 
The quadrupole-hexadecapole 
constrained self-consistent mean-field potential energy surface 
is mapped onto the expectation value of the $sdg$-boson Hamiltonian. 
This procedure determines the parameters 
of the $sdg$-IBM Hamiltonian microscopically. 
Calculated excitation energies and transition strengths are 
compared to the ones obtained with a simpler 
$sd$-IBM, as well as with the experimental data. 
The Gogny-EDF mapped $sdg$-IBM reproduces spectroscopic properties 
of the studied nuclei as reasonably as in the case of the 
previous $sdg$-boson mapping calculations 
that were based on the relativistic EDF, 
indicating that the axial quadrupole-hexadecapole 
method is sound regardless of whether relativistic or 
nonrelativistic EDF is employed. 
The mapped $sdg$-IBM improves 
some of the results in lighter Sm and Gd isotopes 
compared to the mapped $sd$-IBM, 
implying the existence of significant hexadecapole correlations 
in those nuclei. 
For those nuclei with $N \geq 94$, hexadecapole 
effects are minor, and the only significant 
difference between the two boson models 
can be found in the description of $E0$ monopole transitions.
\end{abstract}

\maketitle

\section{Introduction}
The importance of studying deformations in nuclei and their effects on various nuclear properties, e.g. excitation energies and transition strengths, has been recognized for decades \cite{BM,RS}. The most commonly present and extensively studied nuclear deformation is of quadrupole type, and its effects on the positive-parity states are well known. The higher-order correlations which affect the positive-parity states, the hexadecapole correlations, have been significantly less studied, due to their effects often being overshadowed by large quadrupole correlations. Nevertheless, the presence of hexadecapole correlations has been found in a wide range of nuclei \cite{gupta2020, ryssens2023}, and the main effect of such correlations is the appearance of the $4^+$ band in the low-lying excitation energy spectra of some even-even nuclei, with an enhanced $B(E4)$ transition strength from the $4^+$ bandhead state to the $0^+$ ground state. Hexadecapole correlations have also been found to be important in describing certain nuclear reactions and decays \cite{ryssens2023, chi2023, engel2017}. 

The interacting boson model (IBM) \cite{IBM} is one of the most commonly employed  models for studying how nuclear deformations affect  low-lying spectra \cite{casten1988}. The main assumption of the model is that a nucleus can be approximated as a system consisting of a core, represented by the closest doubly-magic nucleus, and the valence nucleons, which form collectives pairs (or bosons). Interactions among valence bosons give rise to the low-lying spectra and transitions. In the simplest version of the model, the $sd$-IBM-1, the nucleons primarily couple into $s$ $(J^{\pi}=0^+)$ and $d$ $(J^{\pi}=2^+)$ bosons, and the neutron and proton bosons are assumed to be identical \cite{IBM}. One particular extension of the simple $sd$-IBM is the inclusion of $g$ bosons $(J^{\pi}=4^+)$, 
allowing for the study of hexadecapole collectivity in nuclei. 
The importance of $g$ bosons in describing low-lying collective 
states has been extensively studied \cite{casten1988, otsuka1981, otsuka1982, otsuka1985, otsuka1988, devi-kota1990, kuyucak1994, vanisacker2010}. 
In recent years, a method has been developed, 
which is to derive the parameters of the IBM Hamiltonian 
from self-consistent mean-field (SCMF) calculations 
based on a microscopic energy density functional (EDF) \cite{nomura2008}. 
The method provides low-energy collective energy 
spectra and reduced transition probabilities, 
and has been successfully applied to study 
quadrupole \cite{nomura2008, nomura2010, nomura2011pt, nomura2012tri}, 
octupole \cite{nomura2013oct, nomura2014}, 
and, recently, hexadecapole correlations \cite{lotina2023, lotina2024}. 
In this paper, we present an extension 
of Refs.~\cite{lotina2023,lotina2024} by studying 
the low-energy quadrupole and hexadecapole collective states of 
even-even rare-earth isotopes $^{150-168}$Sm and $^{152-170}$Gd, 
ranging from transitional to highly deformed nuclei, 
based on the Gogny force \cite{GOGNY1975}. 
The present calculation employs parametrization 
D1S \cite{BERGER1984} of the Gogny interaction, which 
has already been  extensively employed \cite{Robledo2019,hilaire2007}. 
The combination of the Gogny EDFs with the IBM through the 
mapping procedure has been considered and shown to be valid 
to describe a number of nuclear structure phenomena, including 
coexistence of quadrupole shapes (e.g., Refs.~\cite{nomura2016zr,nomura2017ge})
and octupole correlations (e.g., Refs.~\cite{nomura2015,nomura2021oct-u,nomura2021oct-ba,nomura2021oct-zn}).

Recently, Gogny Hartree-Fock-Bogoliubov (HFB) 
calculations, with constraints set on the quadrupole and hexadecapole 
deformation parameters, were performed to compute potential energy surfaces (PESs) in rare-earth nuclei \cite{kumar-robledo2023}. 
In this study, we carry out the mapping procedure 
by using the Gogny-D1S PESs in order to examine 
the effects of the quadrupole-hexadecapole 
coupling on the low-lying spectra of the considered nuclei. 
Since, in previous studies 
\cite{lotina2023, lotina2024}, calculations were performed within the framework of the constrained relativistic mean field (MDC-RMF) model \cite{lu2014, zhao2017}, with the relativistic density-dependent point-coupling (DD-PC1) functional \cite{DDPC1, niksic2011}, this work will allow us to study how the quadrupole-hexadecapole mapping method depends on the choice of the EDF. 
On the other hand, we analyze positive-parity bands in heavy Sm and Gd isotopes  as well as the effects of  hexadecapole collectivity in heavy neutron-rich nuclei, far from the valley of stability. The results of the calculations are compared with the ones obtained with a simpler $sd$-IBM, as well as with the available experimental data \cite{data}.

The paper is organized as follows. 
The theoretical framework, i.e., the Gogny-D1S HFB approach 
and the $sd$-IBM and $sdg$-IBM, is outlined in Sec.~\ref{sec:model}. 
Potential energy surfaces computed  with both 
SCMF and IBM calculations are presented in Sec.~\ref{sec:pes}. 
Spectroscopic properties, including the excitation spectra of low-lying 
states, electric quadrupole, hexadecapole, 
and monopole transition properties, 
are discussed in Sec.~\ref{sec:results}. 
Section~\ref{sec:summary} is devoted 
to the concluding remarks and work perspectives.


\section{Model description\label{sec:model}}

In order to obtain the SCMF PES for the studied nuclei, the HFB equations 
have been solved with constraints on the axial quadrupole $\hat{Q}_{20}$ and hexadecapole $\hat{Q}_{40}$ moment operators \cite{kumar-robledo2023}. The axial quadrupole $\beta_{20} \equiv \beta_2$ and hexadecapole $\beta_{40} \equiv \beta_4$ deformation parameters are related to the expectation values of the multipole moment operators as
 \cite{kumar-robledo2023}
\begin{equation}
    \beta_{\lambda} = \frac{\sqrt{4 \pi (2 \lambda + 1)}}{3 R^{\lambda} A} \left\langle \hat{Q}_{\lambda 0} \right\rangle,
    \label{eq1}
\end{equation}
where $\lambda=2$ or 4, 
$R=1.2 A^{1/3}$ fm and $A$ is the nucleon number. Details on the Gogny-D1S HFB 
calculations can be found in Refs. \cite{BERGER1984, warda2002, DELAROCHE2006}.

In order to calculate low-lying excitation energies and transition strengths, we have
employed
 the $sdg$-IBM-1, referred to simply as $sdg$-IBM. The Hamiltonian is built in the same way as in Ref. \cite{lotina2024}:
\begin{equation}
    \hat{H} = \epsilon_d \hat{n}_d + \epsilon_g \hat{n}_g + \kappa \hat{Q}^{(2)} \cdot \hat{Q}^{(2)} + \kappa(1- \chi^2) \hat{Q}^{(4)} \cdot \hat{Q}^{(4)},
    \label{eq2}
\end{equation}
with $\hat{n}_d= d^{\dagger} \cdot \tilde{d}$ and $\hat{n}_g= g^{\dagger} \cdot \tilde{g}$ being the $d$- and $g$-boson number operators, respectively,
\begin{equation}
\begin{aligned}
  \hat{Q}^{(2)} = &(s^{\dagger} \times \tilde{d} + d^{\dagger} \times s) + \chi  \Big [ \frac{11 \sqrt{10}}{28}(d^{\dagger} \times \tilde{d})^{(2)} \\
      & - \frac{9}{7} \sigma (d^{\dagger} \times \tilde{g} + g^{\dagger} \times \tilde{g})^{(2)} + \frac{3 \sqrt{55}}{14} (g^{\dagger} \times \tilde{g})^{(2)} \Big ] 
\end{aligned}
\label{eq3}
\end{equation}
being the quadrupole operator, and 
\begin{equation}
    \hat{Q}^{(4)} =  s^{\dagger} \times \tilde{g} + g^{\dagger} \times s
    \label{eq4}
\end{equation}
being the hexadecapole operator. This Hamiltonian represents a modified version of the one 
used in Ref. \cite{vanisacker2010}, which satisfies the conditions of three symmetry limits: $[\textnormal{U(6)} \supset \textnormal{U(5)}] \otimes \textnormal{U(9)}$, SU(3) and O(15) \cite{kota1987}. 
Here the quadrupole-quadrupole boson interaction is 
taken to be attractive, $\kappa<0$, 
and a single $g$ boson level is, to a good approximation, above 
that of a $d$ boson, $\epsilon_d < \epsilon_g$. 
In addition, the requirement of the SU(3) symmetry 
limit leads to the constraints that 
$-1 \leq \chi \leq +1$ and $-1 \leq \chi \sigma \leq +1$.

The parameters $\epsilon_d$, $\epsilon_g$, $\kappa$, $\chi$ and $\sigma$ are determined by the mapping procedure \cite{nomura2008}. 
The $sdg$-IBM is connected to the geometric model 
through a coherent state \cite{ginocchio1980}, 
$\ket{\phi} \sim (1+ \tilde{\beta}_2 d^{\dagger}_0 + \tilde{\beta}_4 g^{\dagger}_0)^{N_B} \ket{0}$, with $N_B$ representing the number of bosons, determined by the nearest doubly magic nucleus, and $\ket{0}$ representing the boson vacuum. 
Note that $\tilde{\beta}_2$ and $\tilde{\beta}_4$ 
represent the axial bosonic quadrupole 
and hexadecapole deformation parameters, respectively. 
For Gd and Sm isotopes with $N<106$, the core  is represented by the doubly magic nucleus $^{132}$Sn.  In the case of $^{168}$Sm and $^{170}$Gd, the corresponding core  is $^{176}$Sn since the neutrons occupy the upper 
half of the neutron major shell $N=82-126$.

The parameters of the $sdg$-IBM are fitted 
so that the PES of the IBM, 
$E_{\textnormal{IBM}}(\tilde{\beta}_2, \tilde{\beta}_4)= \bra{\phi} \hat{H} \ket{\phi}/\braket{\phi|\phi}$, should be approximately 
equal to the Gogny-HFB PES in the vicinity of the minimum:
\begin{equation}
    E_{\textnormal{SCMF}}(\beta_2, \beta_4) \approx E_{\textnormal{IBM}}(\tilde{\beta}_2, \tilde{\beta}_4) \; .
    \label{eq5}
\end{equation}
The relation between axial bosonic and fermionic parameters is assumed to be linear, $\tilde{\beta}_2 = C_2 \beta_2$, $\tilde{\beta}_4 = C_4 \beta_4$ as in previous studies with the mapping procedure \cite{nomura2008, nomura2014, lotina2023, lotina2024}. This leaves us with  7 parameters which need to be fitted to reproduce the SCMF PES.

For the $sd$-IBM calculations, 
we have employed the following Hamiltonian \cite{IBM}:
\begin{equation}
    \hat{H}_{sd} = \epsilon_d \hat{n}_d + \kappa \hat{Q}^{(2)} \cdot \hat{Q}^{(2)},
    \label{eq6}
\end{equation}
where the $sd$-IBM quadrupole operator reads
\begin{equation}
    \hat{Q}^{(2)} = s^{\dagger} \tilde{d} + d^{\dagger} s + \chi \left ( d^{\dagger} \times \tilde{d} \right )^{(2)}.
    \label{eq7}
\end{equation}
The parameters $\epsilon_d$, $\kappa$, $\chi$ 
in the $sd$-IBM Hamiltonian are fixed by 
a 1D mapping of the Gogny-D1S PES along the $\beta_4=0$ line:
\begin{equation}
    E_{\textnormal{SCMF}}(\beta_2, \beta_4=0)= E_{sd-\textnormal{IBM}}(\tilde{\beta}_2),
    \label{eq8}
\end{equation}
and the relation between the bosonic and fermionic deformation parameters is assumed to be linear $\tilde{\beta}_2 = C^{sd}_2 \beta_2$.

In order to calculate transition strengths with both the $sdg$-IBM and $sd$-IBM, the transition operators are constructed in the same way as in Refs. \cite{lotina2023, lotina2024}. The quadrupole transition operator is of the form:
\begin{equation}
    \hat{T}(E2)_{sdg/sd} = e_2^{sdg/sd} \hat{Q}^{(2)}_{sdg/sd},
    \label{eq9}
\end{equation}
with  $\hat{Q}^{(2)}_{sdg/sd}$ representing the quadrupole operator of the $sdg$ (\ref{eq3}) or $sd$ (\ref{eq7}) boson model. The hexadecapole operator is chosen, in the case of $sdg$-IBM, to be
\begin{equation}
    \hat{T}^{(E4)}_{sdg} = e_4^{sdg} \left [  s^{\dagger} \tilde{g} + g^{\dagger} s + (d^{\dagger} \times \tilde{d})^{(4)} \right ],
    \label{eq10}
\end{equation}
whereas for $sd$-IBM it is constructed as
\begin{equation}
    \hat{T}^{(E4)}_{sd} = e_4^{sd} (d^{\dagger} \times \tilde{d})^{(4)}.
    \label{eq11}
\end{equation}
The quadrupole and hexadecapole effective charges $e_2^{sdg/sd}$ and $e_4^{sdg/sd}$ are fitted in order to reproduce the measured transition strengths $B(E2; 2_1^+ \rightarrow 0_1^+)$ and $B(E4; 4_1^+ \rightarrow 0_1^+)$, from the first $2^+$ and $4^+$ states to the ground state $0^+$, respectively. The monopole transition operator  \cite{vanisacker2012} reads
\begin{equation}
    \hat{T}^{(E0)}_{sdg} = (e_n N + e_p Z) \left ( \eta \frac{\hat{n}_d}{N_B} + \gamma \frac{\hat{n}_g}{N_B} \right ),
    \label{eq12}
\end{equation}
for the $sdg$-IBM, and
\begin{equation}
    \hat{T}^{(E0)}_{sd} = (e_n N + e_p Z) \eta \frac{\hat{n}_d}{N_B},
\end{equation}
for the $sd$-IBM.

Following Ref. \cite{vanisacker2012}, we set neutron and proton effective charges to the values 
$e_p=e$ and $e_n=0.5e$, while the values of parameters $\eta$ and $\gamma$ are taken to be $\eta=\gamma=0.75$ fm$^2$. 
Experimental data for energies and transition strengths have been taken 
from the NNDC database \cite{data}.

%
%
\begin{figure*}    
\begin{center}
\includegraphics[width =\textwidth]{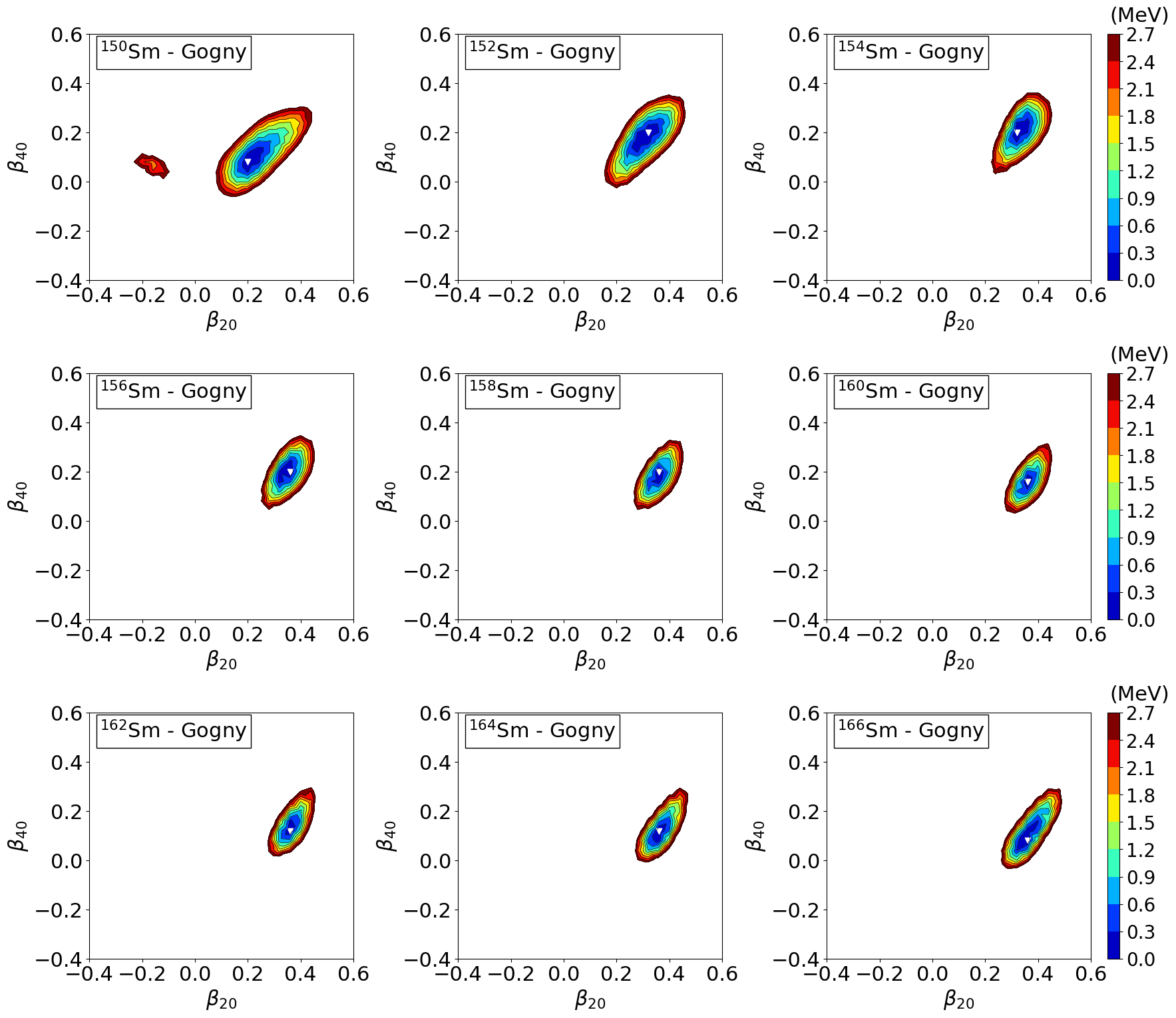}
\caption{Axially symmetric quadrupole ($\beta_{20}$) and 
hexadecapole ($\beta_{40}$) 
constrained potential energy surfaces for  
$^{150-166}$Sm.
The energy difference between neighbouring contours is 
0.1 MeV. The absolute minimum is indicated by 
an open triangle. Results have been obtained with the Gogny-D1S energy density 
functional.} 
\label{SMGOGNY}
\end{center}
\end{figure*}

\begin{figure*}    
\begin{center}
\includegraphics[width = \textwidth]{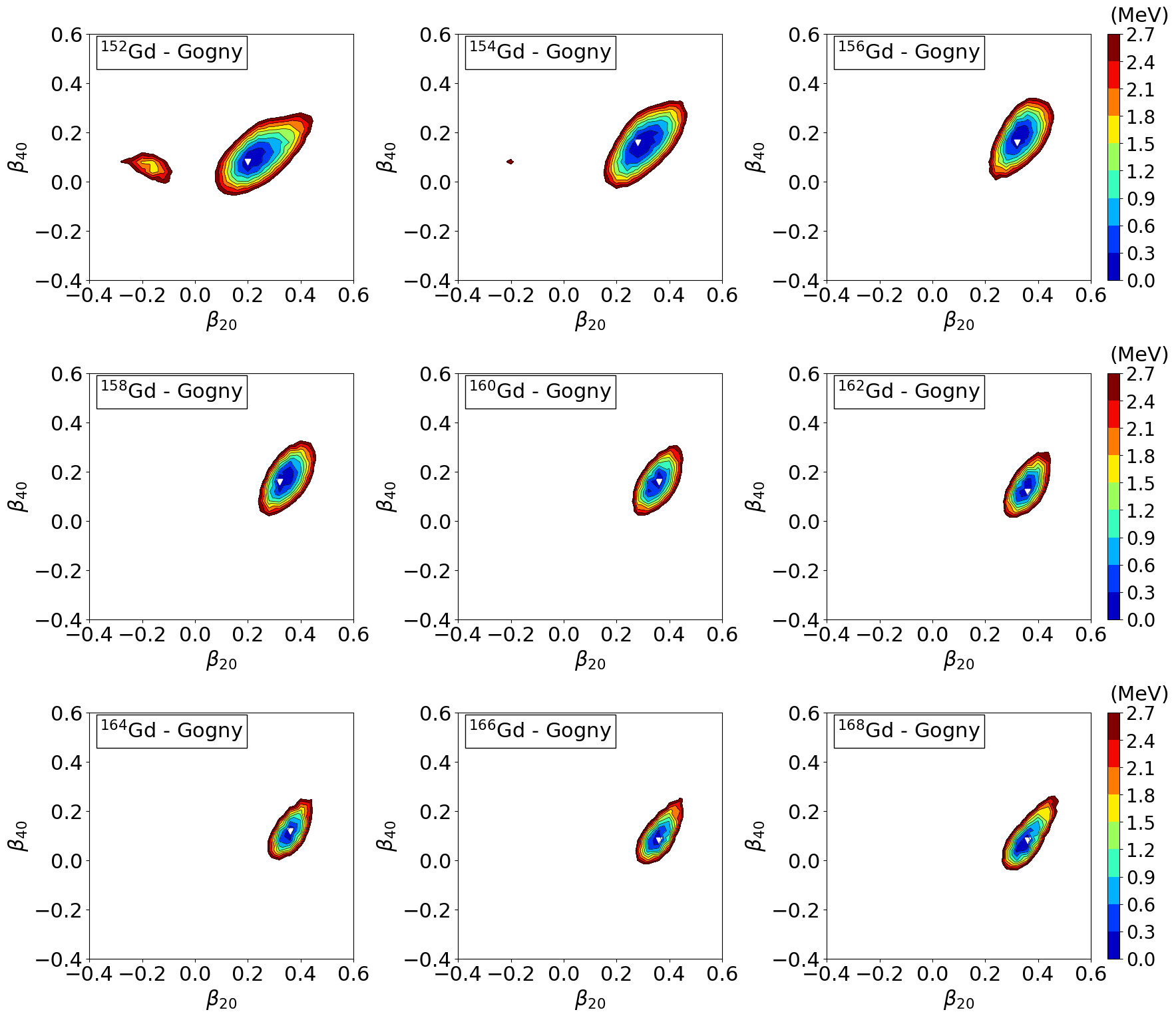}
\caption{The same as in  Fig. \ref{SMGOGNY}, but for 
$^{152-168}$Gd.
} 
\label{GDGOGNY}
\end{center}
\end{figure*}

\begin{figure*}    
\begin{center}
\includegraphics[width =\textwidth]{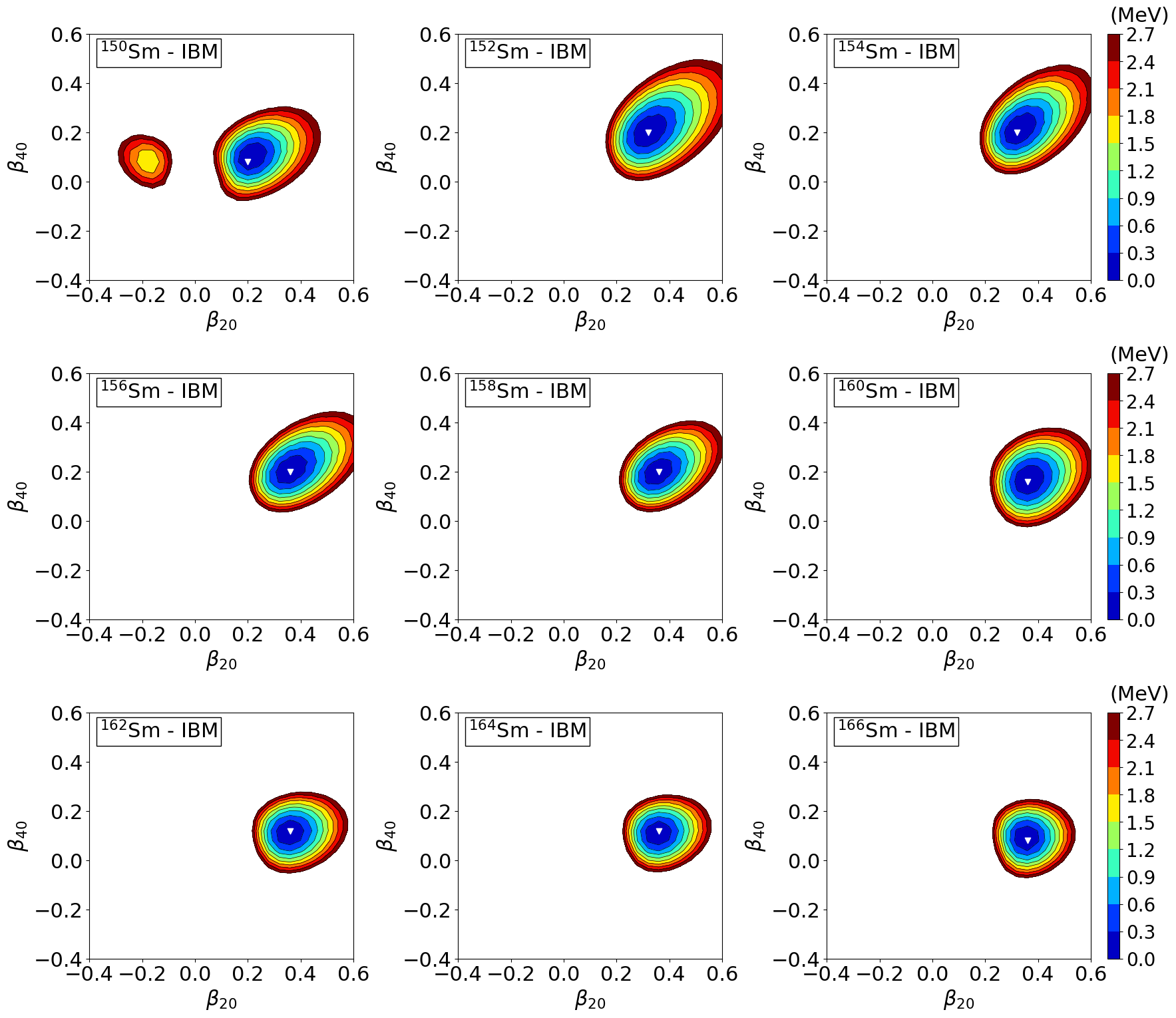}
\caption{The same as in  Fig. \ref{SMGOGNY}, but for the mapped $sdg$-IBM potential energy surfaces of 
$^{150-166}$Sm.
} 
\label{SMIBM}
\end{center}
\end{figure*}

\begin{figure*}    
\begin{center}
\includegraphics[width =\textwidth]{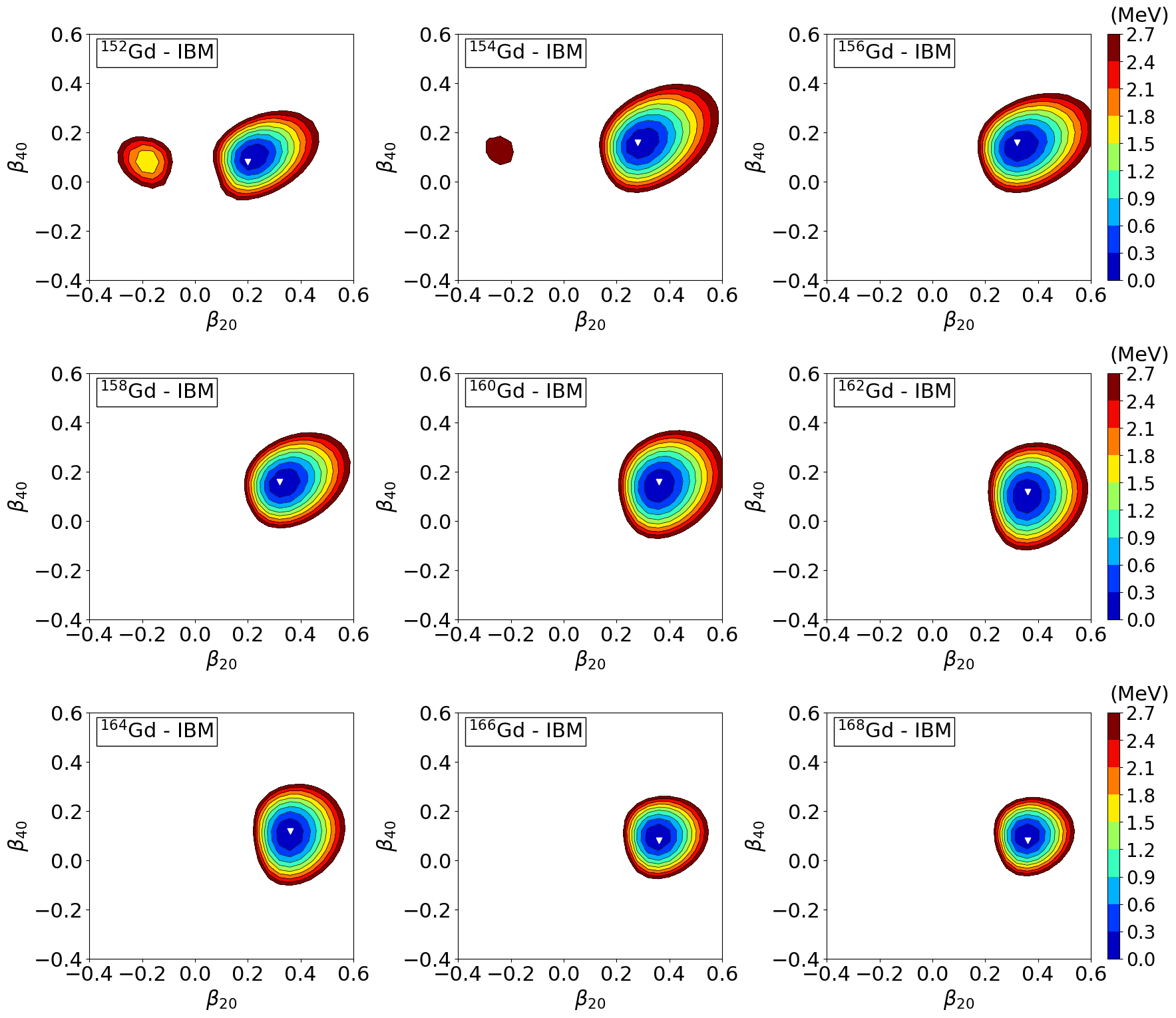}
\caption{The same as in  Fig. \ref{SMGOGNY}, but for the mapped $sdg$-IBM potential energy surfaces of 
$^{152-168}$Gd.
} 
\label{GDIBM}
\end{center}
\end{figure*}

\section{Mapping the SCMF results onto the IBM space\label{sec:pes}}

Figures \ref{SMGOGNY} and \ref{GDGOGNY} show the Gogny-D1S PESs 
obtained for even-even Sm and Gd isotopes 
with neutron numbers within the range 
$N=88-106$, up to an energy of 2.7 MeV above the HFB ground state. 
The PESs for the $N=106$ isotones are not shown 
due to their similarity to those 
of the $N=104$ ones. As can be seen from the figures, the quadrupole and hexadecapole deformations 
associated with the absolute minima of the PESs start from the values $\beta_2^{\textnormal{min}}=0.2$ and $\beta_{4}^{\textnormal{min}}=0.08$, respectively. They increase with neutron number until $N=94$, after which 
 hexadecapole deformations start decreasing, dropping to  $\beta_{4}^{\textnormal{min}}=0.04$ at $N=106$. 
Similar to previous (mapped) IBM calculations, based  on the relativistic DD-PC1 
functional \cite{lotina2023, lotina2024}, Sm isotopes exhibit larger HFB ground state $\beta_4$ values as  
compared to Gd isotopes. On the other hand, the ground state $\beta_2$ deformations increase up to $N=94$.
For larger neutron numbers, those quadrupole deformations decrease, reaching the value
$\beta_{2}^{\textnormal{min}}=0.32$ for 
 $N=106$ isotopes. An important thing to observe is that the PESs appear to be "tilted" with respect to $\beta_{20}$ and $\beta_{40}$ axes, indicating a strong coupling between the axial quadrupole and hexadecapole degrees of freedom, which has already been discussed in Ref. \cite{kumar-robledo2023}. Note, that the results obtained for $N=88-96$ isotopes agree well with the ones 
 obtained with the DD-PC1 EDF \cite{lotina2023, lotina2024}.

The corresponding (mapped)  $sdg$-IBM PESs are depicted
in Figs.~\ref{SMIBM} and \ref{GDIBM}. The mapping procedure reproduces
basic features of the fermionic PESs 
such as the position of the absolute minima, the oblate saddle points 
for $N=88$ nuclei as well as the overall shape of the surface. The strong quadrupole-hexadecapole coupling is well reproduced in $N \leq 100$ isotopes, whereas in $N > 100$ isotopes, the coupling is predicted to be less strong.
Note that the IBM PESs are significantly flatter 
than the HFB ones, due to the restricted boson (valence) space 
as compared to the Gogny-HFB model space. 
This also explains why the mapping procedure 
is not able to reproduce the ``elongated'' 
shapes of the PESs for $N \geq 100$ nuclei. 
The 1D $sd$-IBM mapping has also been carried out in all the studied isotopes to reproduce some of the properties of the $E_{\textnormal{SCMF}}(\beta_2, \beta_4=0)$ curve such as the position of the absolute minimum, the position and 
energy of the oblate saddle point, and the energy of the spherical $\beta_2=0$ configuration.

\begin{figure*}    
\begin{center}
\includegraphics[width = \linewidth]{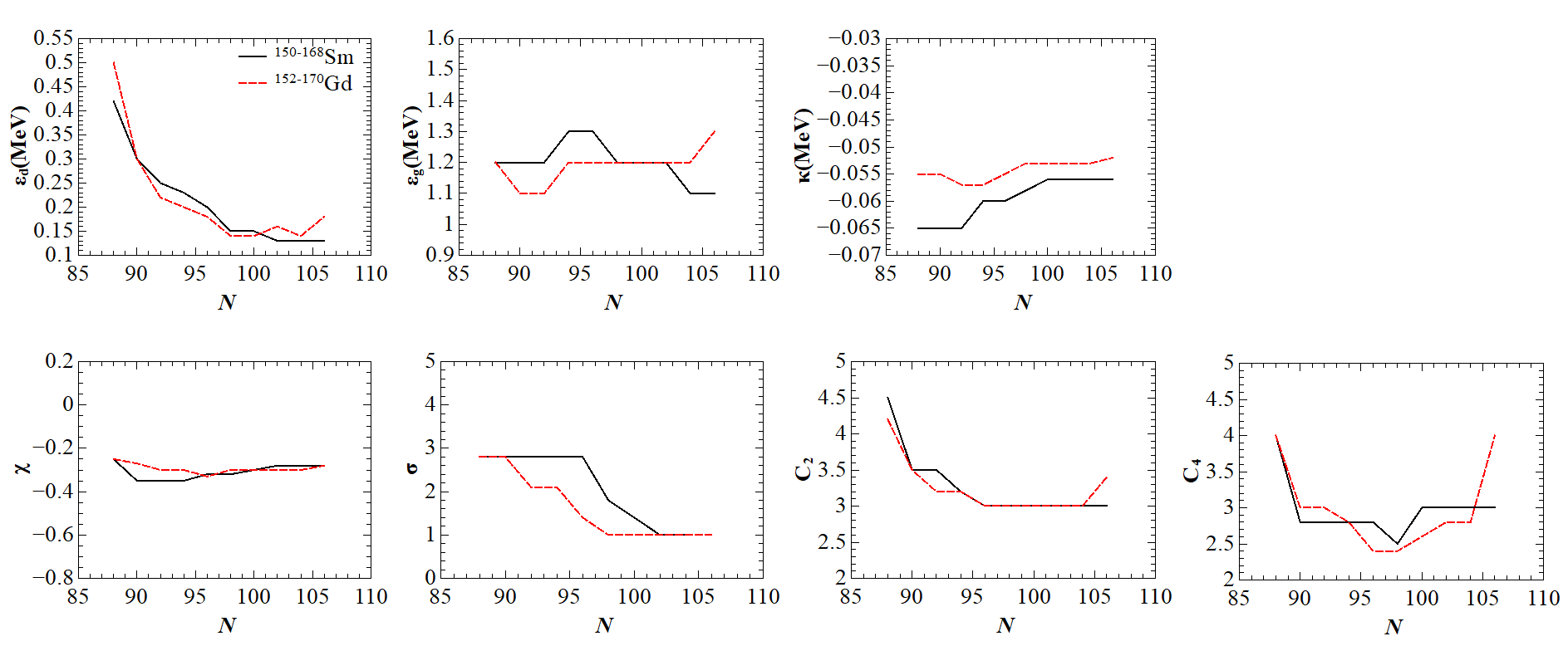}
\caption{Parameters of the $sdg$ - IBM Hamiltonian Eq.(\ref{eq2}), as functions of the neutron number $N$.} 
\label{sdgpar}
\end{center}
\end{figure*}

\begin{figure*}    
\begin{center}
\includegraphics[width =0.6\linewidth]{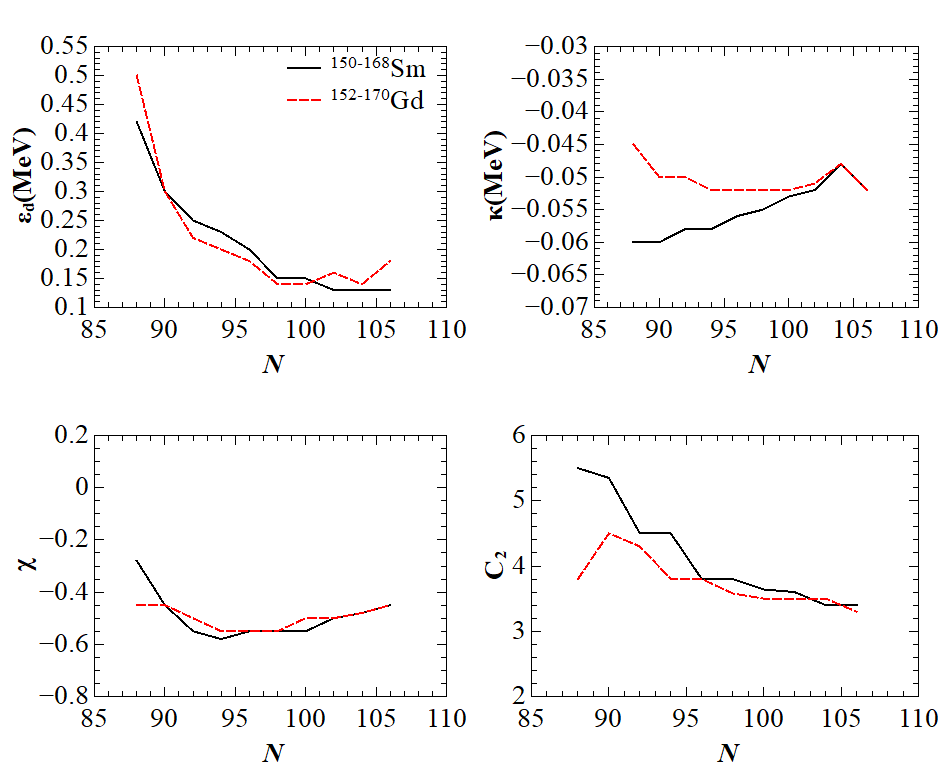}
\caption{Parameters of the $sd$ - IBM Hamiltonian Eq.(\ref{eq6}), as functions of the neutron number $N$.} 
\label{sdpar}
\end{center}
\end{figure*}

The $sdg$- and $sd$-IBM parameters are shown in Figs. \ref{sdgpar} and \ref{sdpar}, as functions of 
the neutron number $N$. The $d$-boson energy $\epsilon_d$ has a very similar behavior in both models. It 
decreases up to $N=100$, after which it stops decreasing, and even increases at $N=106$, which corresponds to 
lower $\beta_2^{\rm min}$ values. Moreover, the parameters $\kappa$, $\chi$ and $C_2$ also display a similar behavior 
in both models, whereas in the $sdg$-IBM the $\chi$ parameter is characterized by significantly lower absolute values, compared to the $sd$-IBM ones. Regarding those parameters appearing only in the $sdg$-IBM, the $g$-boson energy $\epsilon_g$ takes values between $\epsilon_g=1.1$ and $\epsilon_g=1.3$ MeV. The $C_4$ parameter decreases until $N=96$, after which it starts to increase, which corresponds to $\beta_4^{\rm min}$ values becoming lower for $N \geq 96$ isotopes. 
Finally, with increasing neutron number, the parameter $\sigma$, which accounts for the quadrupole-hexadecapole coupling in the $sdg$-IBM, 
drops from $\sigma=2.8$ to $\sigma=1.0$, which also corresponds to a decrease in $\beta_4^{\textnormal{min}}$. This explains why the coupling is predicted to be less strong in $N > 100$ isotopes by the $sdg$-IBM.
It should also be noted that $\sigma$ decreases faster in Gd isotopes, compared to the Sm ones, which can be explained from the fact that $\beta_4^{\textnormal{min}}$ values are smaller in Gd isotopes with the same $N$.

%
%
 
\section{Results of the spectroscopic calculations\label{sec:results}}

In this section, we discuss the excitation energies 
and transition strengths obtained from the diagonalization 
of the IBM Hamiltonian, using 
the computer program ARBMODEL \cite{arbmodel}. 
The results of the $sdg$-IBM and $sd$-IBM will be 
compared to illustrate the effects of $g$ bosons.
The results obtained from both models will  also be 
compared with the available experimental data 
at NNDC \cite{data}.

\subsection{Excitation energies}

\begin{figure}
\begin{center}
\includegraphics[width =\linewidth]{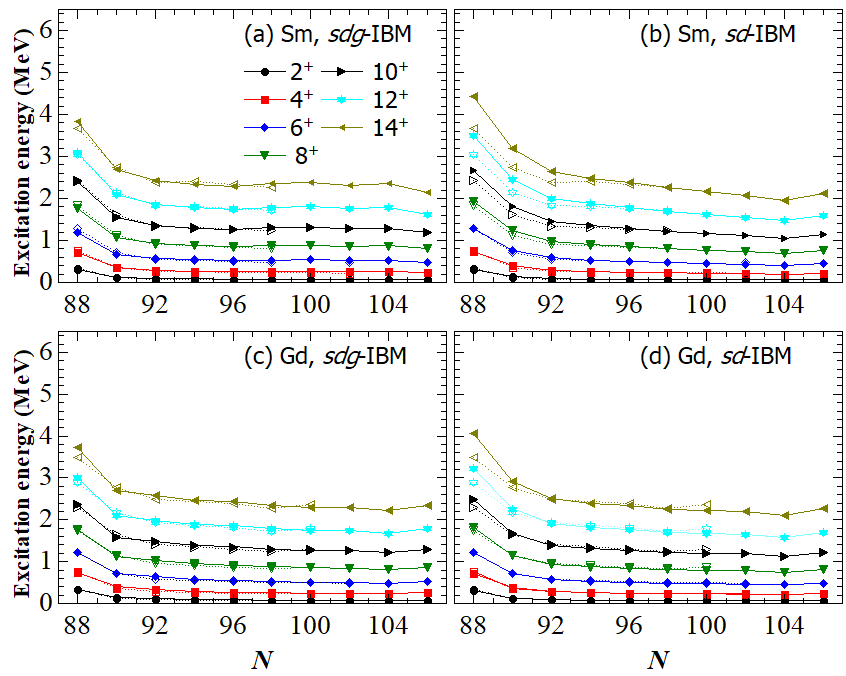}
\caption{Excitation energies of the yrast band 
states up to spin $J^{\pi}=14^+$ as functions 
of the neutron number $N$ within 
the mapped $sdg$-IBM (left column) and $sd$-IBM (right column), 
represented by solid symbols connected by solid lines. 
Experimental data are taken from Ref.~\cite{data}, 
and are depicted as open symbols connected by 
dotted lines.}
\label{yrast}
\end{center}
\end{figure}

Figure \ref{yrast} shows the  excitation 
energies corresponding to the yrast band states with even spin 
$J^{\pi}=2^+$ to $14^+$. 
As expected, the $sdg$-IBM improves the description of the 
excitation energies of higher-spin states 
with $J^{\pi} = 12^+$ and $14^+$ in transitional $N=88$ nuclei. 
For Sm nuclei, the $sdg$-IBM gives a better description of the 
energy levels up to $N=92$ than the $sd$-IBM. 
This can be explained by the fact that large amounts of 
$g$-boson contributions are present in the higher-spin states, 
as indeed, the expectation 
value of the $g$-boson number operator $\braket{\hat{n}_g}$ 
is greater than or equal to 1 in those isotopes with $N\leq 92$. 
Note that for Sm isotopes with $N \leq 92$, 
the $\beta_4^{\rm min}$ values are larger compared to those for Gd isotopes with the same $N$. For the $N > 92$ isotopes, there is no significant difference between both models concerning the description of the ground state bands. 
This is not surprising, since as the number of valence nucleons 
increases for deformed nuclei even the $sd$-IBM is good enough 
to provide an accurate description of the yrast-band levels. 
In any case, in general the $sdg$-IBM does provide 
some significant improvements in the description 
of the yrast band, compared to the simpler $sd$-IBM.

\begin{figure}
\begin{center}
\includegraphics[width=\linewidth]{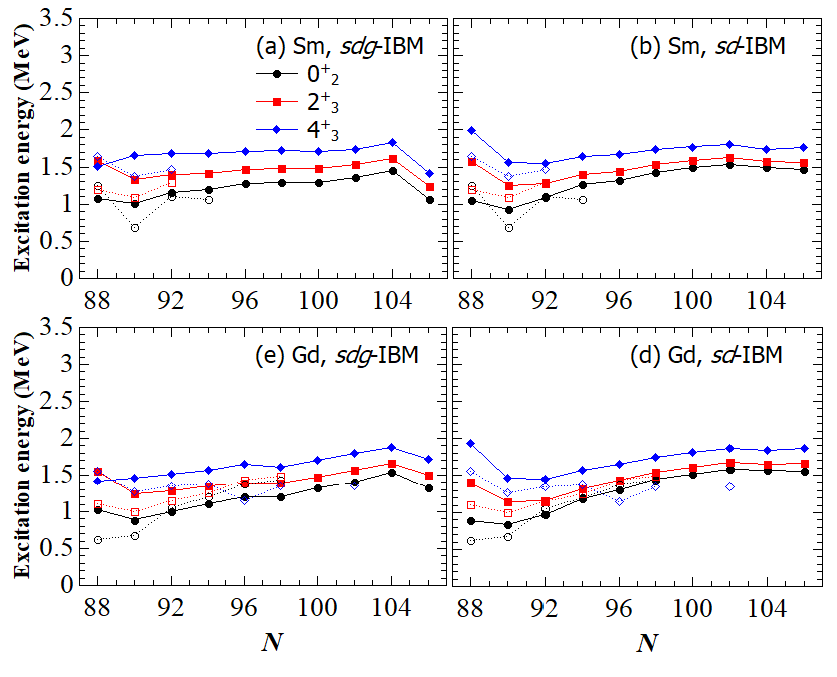}
\caption{The same as in Fig.~\ref{yrast}, but for 
the excitation energies of $0_{2}^+$, $2_{3}^+$, and $4_{3}^+$ states.} 
\label{0+}
\end{center}
\end{figure}

Figure~\ref{0+} depicts the excitation energies 
of the $0_{2}^+$, $2_{3}^+$ and $4_{3}^+$ states, 
which are associated with 
the $K^{\pi}=0^+$ band in the present study. 
The $sdg$-IBM predicts a significantly low-lying $4_3^+$ 
level for $N=88$ isotopes. 
This agrees better with the experimental data than 
in the corresponding  $sd$-IBM results. 
For other isotopes, there is no significant difference between the two models, as they both predict similar energies for all three states. Overall, the $sdg$-IBM does not improve the description of the $0^+$ band significantly. 

\begin{figure}
\begin{center}
\includegraphics[width=\linewidth]{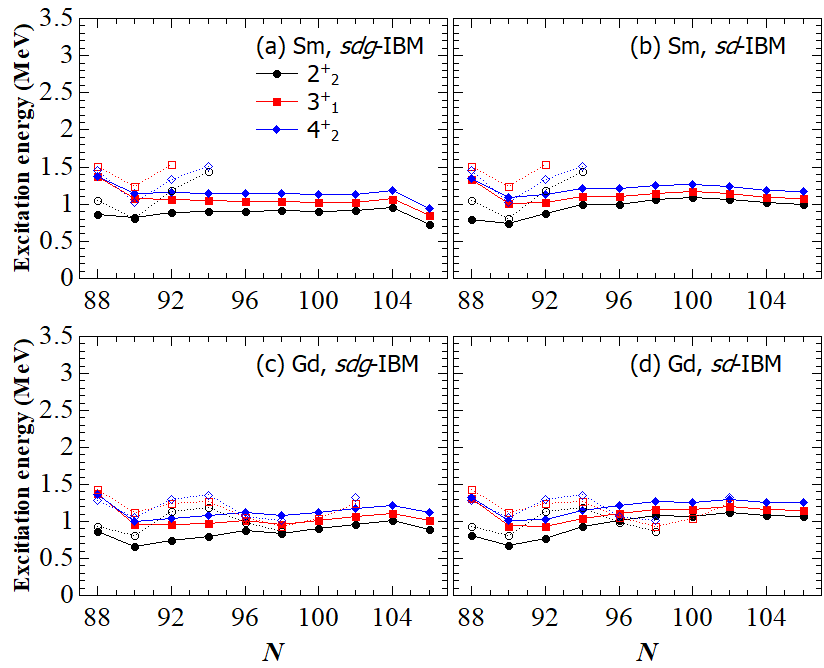}
\caption{The same as in Fig.~\ref{yrast}, but for the excitation energies of
$2_{2}^+$, $3_{1}^+$, and $4_{2}^+$ states.} 
\label{gamma}
\end{center}
\end{figure}

Figure \ref{gamma} shows the excitation energies of the $2^+_2$, $3^+_1$, and $4^+_2$ states, which can be identified as members of the $\gamma$-vibrational band in the present calculation. 
Both models predict similar energies for Sm and Gd isotopes. 
While the calculated energies of the $\gamma$-vibrational band states are in good agreement with the experimental data for the nuclei with $N \leq 90$ and $N \geq 96$, both models underestimate the energies for those with $N=92$ and 94. 
Overall, the presence of $g$ bosons in the IBM calculations does not seem to have any significant effects on the $\gamma$-vibrational band.
The description of the $\gamma$-vibrational band could be improved by including both three-body terms \cite{nomura2012tri} and the  $\gamma$ degree of freedom at the HFB level. However, such an extension 
is beyond the scope of this paper. 
Work along these lines will be reported in future publications.

%
%

\subsection{Transition strengths}
\subsubsection{$E2$ transitions}

\begin{figure}
\begin{center}
\includegraphics[width=\linewidth]{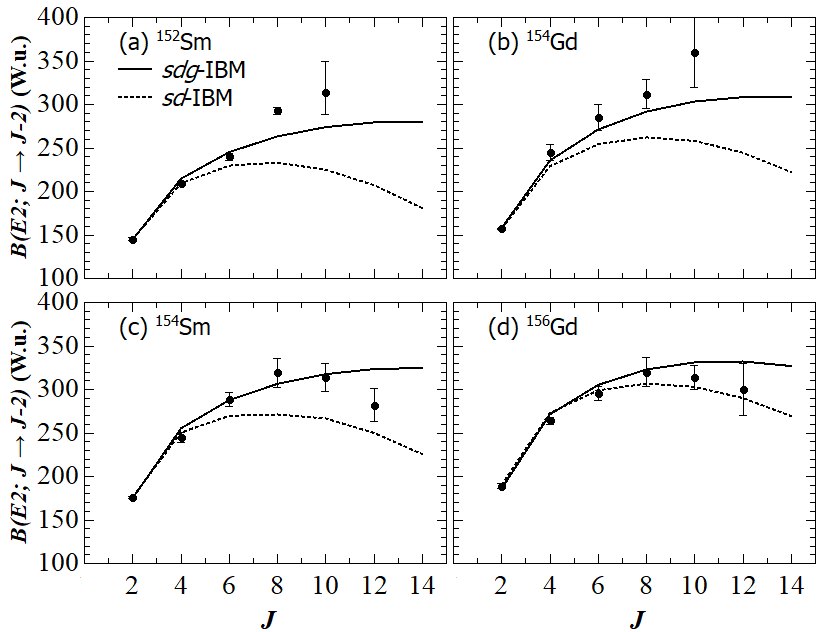}
\caption{$B(E2)$ transition strengths 
in the ground state band of the well-deformed 
$N=90$ (first row) and $N=92$ (second row) isotopes 
as functions of spin $J$, 
computed with the mapped $sdg$-IBM 
(solid curves) and $sd$-IBM (dotted curves). 
The experimental data, represented by solid circles, 
are taken from Ref.~\cite{data}.} 
\label{E2}
\end{center}
\end{figure}

Figure \ref{E2} shows the $B(E2; J \rightarrow J-2)$ 
transition strengths in the ground state bands of the 
well deformed $N=90$ and 92 isotones. 
These nuclei are specifically considered, 
since there are experimental data on $E2$ transitions. 
In Sm isotopes, the $sdg$-IBM significantly improves the description of $B(E2)$ transition strengths for $J^{\pi} \geq 6^+$. However, for $^{154}$Gd, the $sdg$-IBM predicts $B(E2)$ strengths significantly larger 
than the $sd$-IBM, which are however still smaller 
than the measured values. 
The discrepancy points towards some missing correlations 
other than the hexadecapole ones, 
that could also contribute to the $B(E2)$ strengths. 
Due to the large experimental error bars, in $^{156}$Gd 
it is not possible to conclude whether $sdg$-IBM improves the description of the $E2$ transitions in the yrast 
band or not. 
The fact that the $sdg$-IBM predicts larger $B(E2)$ transition 
strengths than the $sd$-IBM can be attributed to 
the large values of the parameter $\sigma$ (see Fig. \ref{sdgpar}). 
In the case of  
$^{152, 154}$Sm and $^{154}$Gd the value $\sigma=2.8$ leads to a significant contribution to the 
$B(E2)$ transition strengths arising from the $\left [ d^{\dagger} \times \tilde{g} + g^{\dagger} \times \tilde{d} \right ]^{(2)}$ part of the quadrupole $\hat{Q}^{(2)}$ operator of Eq.(\ref{eq3}). Moreover, the 
$\sigma=2.1$ value obtained for $^{156}$Gd leads to similar $sdg$-IBM and $sd$-IBM $B(E2)$ strengths. 
Note that the $\sigma$ values for $^{154,156}$Gd 
are larger than those previously obtained \cite{lotina2024} using 
microscopic input from the relativistic DD-PC1 EDF.

%
%

\subsubsection{Hexadecapole transitions}

\begin{figure}
\begin{center}
\includegraphics[width=\linewidth]{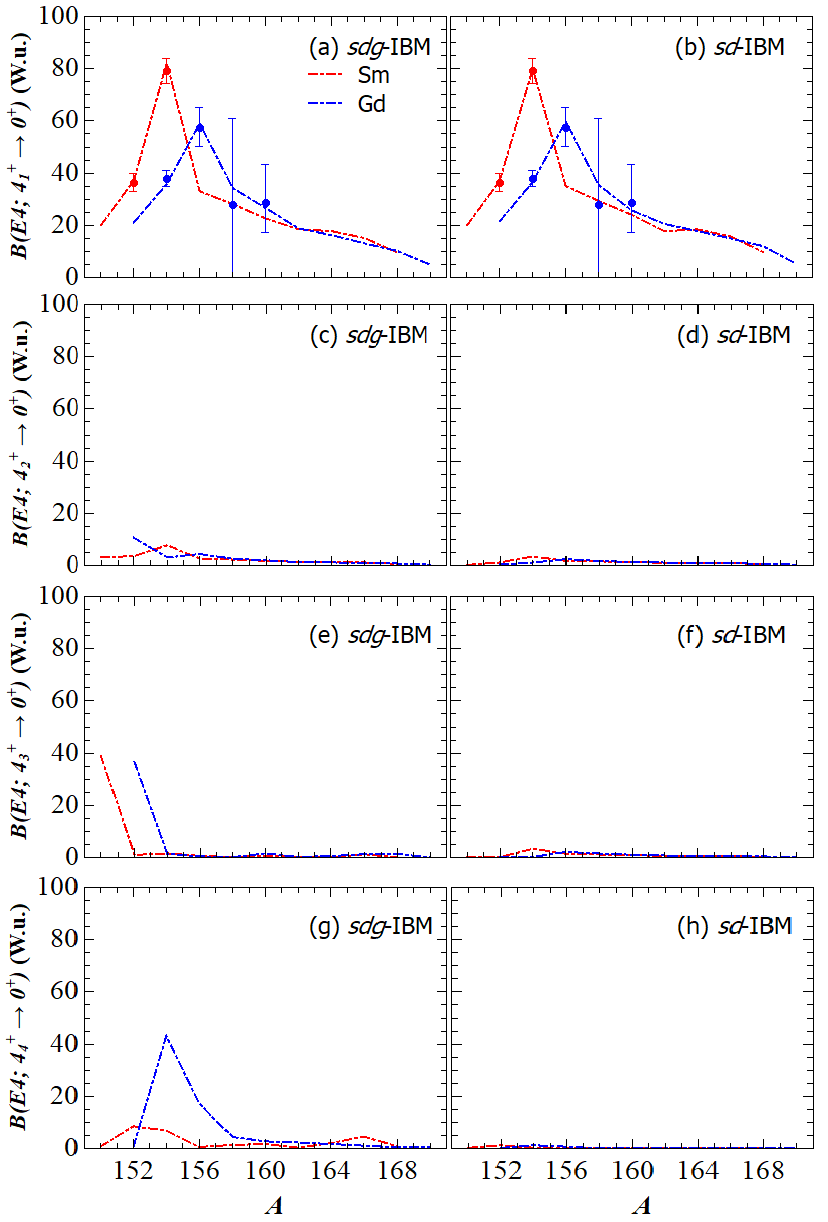}
\caption{$B(E4; 4^+_n \to 0^+_1)$ ($n=1,2,3,4$) 
transition strengths as functions 
of the mass number $A$, computed with 
the mapped $sdg$-IBM (left column) 
and $sd$-IBM (right column). 
Experimental data, indicated by solid circles in the plots, are taken from 
Refs.~\cite{data, ronningen1977, wollersheim1977, ronningen21977}.} 
\label{E4}
\end{center}
\end{figure}

The $B(E4; 4_n^+ \rightarrow 0_1^+)$ ($n=1,2,3,4$) reduced 
transition probabilities obtained in the mapped calculations 
are plotted in Fig.~\ref{E4}, as functions of the nucleon number $A$
in order to avoid overlapping between the data for Sm and Gd isotopes. The 
$e_4^{sdg/sd}$ effective charges are fitted 
to the experimental  $B(E4; 4_1^+ \rightarrow 0_1^+)$ values 
\cite{data, ronningen1977, wollersheim1977, ronningen21977}. Results 
for $B(E4; 4_1^+ \rightarrow 0_1^+)$ strengths are plotted in 
Fig.~\ref{E4}(a) and \ref{E4}(b), while $B(E4)$ values from non-yrast $4_n^+$ (with $n=2,3,4$)  to the $0^+$ ground 
states are shown in Figs.~\ref{E4}(c)-\ref{E4}(h). 
The $4_2^+$ state typically belongs to the $\gamma$-vibrational 
band, where $g$ bosons do not play a significant role. 
Thus both models predict weak $E4$ transitions
in this case. 
Large $sdg$-IBM $B(E4)$ values are obtained for the  
$4_3^+$ states in $^{150}$Sm and $^{152}$Gd. 
The strong $E4$ transition, predicted 
for the $4_4^+$ state in $^{154}$Gd, agrees well with 
the experimental identification 
of this state as a bandhead of the $K=4^+$ band \cite{data}. 
On the other hand, the $sd$-IBM gives 
negligibly small $B(E4)$ reduced transition probabilities. 
This indicates a significant difference 
between the $sdg$-IBM and the simpler $sd$-IBM. 
Nevertheless, experimental data on $B(E4)$ strengths 
in rare-earth nuclei are still required to understand whether the $sdg$-IBM describes them better.

%
%

\subsubsection{Monopole transitions}

\begin{figure}
\begin{center}
\includegraphics[width=\linewidth]{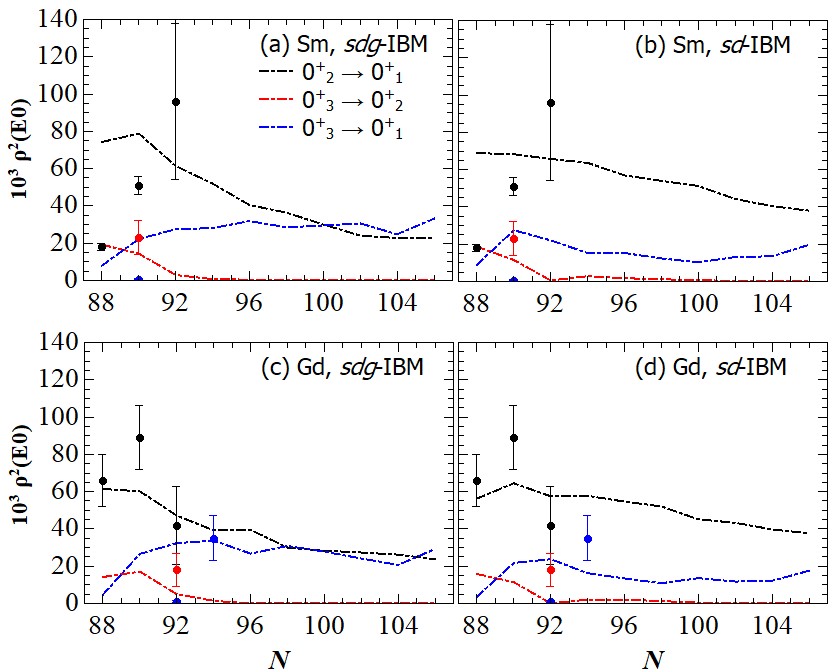}
\caption{$\rho^{2}(E0; 0_i^+ \rightarrow 0_j^+)$ 
values as functions of the neutron number $N$ 
for Sm and Gd isotopes, computed with 
the mapped $sdg$-IBM (left column) 
and $sd$-IBM (right column). 
Experimental values, plotted as solid circles, are taken from 
Refs.~\cite{data,kibedi2005}.} 
\label{E0}
\end{center}
\end{figure}

Figure~\ref{E0} shows the monopole strengths $\rho^{2}(E0; 0_i^+ \rightarrow 0_j^+)$, with $i=2,3$ and $j=1,2$, as functions of the neutron number $N$. In $N \leq 92$ isotopes, both models yield similar values of monopole strengths. As can be 
seen from the figure, $\rho^{2}(E0; 0_3^+ \rightarrow 0_2^+)$ vanishes for $N >92$.  The  
$sdg$-IBM predicts  a much sharper decrease (increase) of  $\rho^{2}(E0; 0_2^+ \rightarrow 0_1^+)$
[$\rho^{2}(E0; 0_3^+ \rightarrow 0_1^+)$], with increasing neutron number, than the $sd$-IBM.

One can see from Figs.~\ref{E0}(a) and (c) that 
for those Sm and Gd nuclei with 
$N \geq 92$, the $sdg$-IBM predicts the monopole 
strengths $\rho^{2}(E0; 0_2^+ \rightarrow 0_1^+)$ 
and $\rho^{2}(E0; 0_3^+ \rightarrow 0_1^+)$ to be more or 
less similar to each other, whereas in the $sd$-IBM these 
monopole transition strengths are quite at variance, 
that is, the $\rho^{2}(E0; 0_2^+ \rightarrow 0_1^+)$ are larger than 
the $\rho^{2}(E0; 0_3^+ \rightarrow 0_1^+)$ by a factor of 2 to 3. 
In addition, the $\rho^{2}(E0; 0_2^+ \rightarrow 0_1^+)$ values 
obtained from the mapped $sdg$-IBM are smaller than 
those from the $sd$-IBM for neutron-rich nuclei with $N \geq 94$. 
The reduction of the $0_2^+ \rightarrow 0_1^+$ $E0$ transition 
in the case of the $sdg$-IBM has been already been 
discussed in Ref.~\cite{vanisacker2012}. 
On the other hand, the $sdg$-IBM provides larger 
$\rho^{2}(E0; 0_3^+ \rightarrow 0_1^+)$ values for the 
$N\geq 94$ nuclei than the $sd$-IBM. 
These quantitative differences in the predicted $\rho^2(E0)$ 
values between the two sets of the IBM calculations 
represent the most significant difference between 
the $sdg$-IBM and $sd$-IBM in the neutron-rich Sm and Gd isotopes 
with $N \geq 94$. 
As in the case of the hexadecapole transitions, experimental information about the monopole transitions is limited. It is therefore not possible to tell whether the $sdg$-IBM represents an improvement over the simpler $sd$-IBM in this respect.

\section{Summary\label{sec:summary}}

We have presented an extensive analysis of the quadrupole-hexadecapole collectivity in even-even rare-earth isotopes $^{150-168}$Sm and $^{152-170}$Gd, and the effects of hexadecapole correlations on low-lying excitation energies and transition strengths. The calculations were performed in the framework of the Gogny D1S EDF and the $sdg$-IBM and $sd$-IBM. The calculated results for the excitation spectra were shown to be in good agreement with the experimental data at the same level of accuracy as our previous mapped $sdg$-IBM studies based on the relativistic EDF \cite{lotina2023,lotina2024}. We have thus found that the quadrupole-hexadecapole mapping method does not significantly depend on the choice of the EDF.

The mapped $sdg$-IBM improves the description of the high-spin yrast states in the nuclei with $N=88$, as well as in the Sm isotopes with $N=90$ and 92. In heavier rare-earth nuclei, the contribution of $g$ bosons to the ground state band is insignificant.
In the case of the excited $K=0^+$ and $\gamma$-vibrational bands, the $sdg$-IBM results did not differ significantly from the $sd$-IBM ones, with only some minor improvements of the $4_3^+$ states of the $K=0^+$ band at $N=88$ in the $sdg$-IBM. 
In well-deformed $N=90$ and 92 nuclei, the $sdg$-IBM predicts stronger $B(E2)$ transition strengths for the yrast states with spin $J^{\pi} \geq 6^+$, which is in good agreement with the experimental data. The effect is more visible in $^{152,154}$Sm, for which pronounced hexadecapole mean-field minima were found in the PESs. The choice of the Gogny D1S EDF seems to lead to somewhat stronger yrast band $B(E2)$ values compared to the ones obtained with the relativistic DD-PC1 EDF.
The existence of the $K=4^+$ band with an enhanced $B(E4; 4^+ \rightarrow 0^+)$ transition to the ground state is predicted by the $sdg$-IBM in $^{150}$Sm and $^{152,154}$Gd. In heavier isotopes, such bands are also predicted; however, the bandhead in those nuclei is predicted to be a higher-lying $4_{n \geq 5}^+$ state.
Regarding monopole strengths, the $sdg$-IBM predicts lower values of $\rho^{2}(E0; 0_2^+ \rightarrow 0_1^+)$ and larger values of $\rho^{2}(E0; 0_3^+ \rightarrow 0_1^+)$ in neutron-rich isotopes with $N\geq 94$ than the $sd$-IBM. This is in good agreement with previous theoretical calculations, and represents the most pronounced hexadecapole correlation effect on the low-lying spectra in very neutron-rich Sm and Gd isotopes. Due to the lack of experimental data on such transitions, it remains to be seen whether the $sdg$-IBM correctly predicts the behavior of monopole transitions in those isotopes.

Now that the mapping method has been shown to be adequate for describing the quadrupole-hexadecapole collectivity in nuclei, regardless of the choice of the EDF, the method can be extended to include even-odd and odd-odd nuclei. The method should also be applied to heavier rare-earth nuclei up to W and Os nuclei, where more experimental data on $E4$ transitions are available, and to explore possible hexadecapole correlations in other regions of the nuclear chart such as actinides. The method could also be extended to include proton and neutron degrees of freedom (i.e., $sdg$-IBM-2), which become relevant to describe phenomena like mixed-symmetry states and scissor modes.

\acknowledgments
Most of the work by L.L. was done at the University of Zagreb and was funded within 
the Tenure Track Pilot Programme of 
the Croatian Science Foundation and 
the \'Ecole Polytechnique F\'ed\'erale de Lausanne, 
and the project TTP-2018-07-3554 
Exotic Nuclear Structure and Dynamics, 
with funds from the Croatian-Swiss Research Programme. This work has also been supported by the Spanish Agencia Estatal de Investigaci\'on (AEI) of the Ministry of Science and Innovation (MCIN) under Grant Agreement No. PID2021-127890NB-I00. L.L. acknowledges support by the ``Ram\'on y Cajal'' Grant No. RYC2021-031880-I funded by MCIN/AEI/10.13039/501100011033 and the European Union-''NextGenerationEU''. The work of R.R.-G. was partially 
supported through Grant No. PID2022-136228NB-C22 funded by 
MCIN/AEI/10.13039/501100011033/FEDER, UE and "ERDF A way of making 
Europe."

\bibliography{refs}

\end{document}